\documentclass[ preprint,
                superscriptaddress, 
                aps, 
                prx, 
                citeautoscript]{revtex4-2}
\usepackage{graphicx}
\usepackage{dcolumn}
\usepackage{bm}
\usepackage{amsmath,amssymb}
\usepackage[version=4]{mhchem}
\usepackage{xcolor}
\usepackage{siunitx}
\DeclareSIUnit\angstrom{\text {Å}}
\usepackage{braket}
\usepackage{xr}
\usepackage{float}
\usepackage{multirow}
\usepackage{wrapfig}
\usepackage{gensymb}
\usepackage{quotes}
\usepackage{xr}
\begin{document}

\title{Fine-Tuning Exciton Polaron Characteristics via Lattice Engineering in 2D Hybrid Perovskites}
\author{Katherine~A~Koch}
\affiliation{Department of Physics \& Center for Functional Materials, Wake Forest University, Winston-Salem, North Carolina 27109, United~States}
\author{Martin~Gomez-Dominguez}
\affiliation{School of Materials Science and Engineering, Georgia Institute of Technology, 771 Ferst Dr NW, Atlanta, GA~30332, United~States}
\author{Esteban~Rojas-Gatjens}
\affiliation{School of Chemistry and Biochemistry, Georgia Institute of Technology, 771 Ferst Dr NW, Atlanta, GA~30332, United~States}
\author{Alexander~Evju}
\affiliation{Department of Physics \& Center for Functional Materials, Wake Forest University, Winston-Salem, North Carolina 27109, United~States}
\author{K~Burak~Ucer}
\affiliation{Department of Physics \& Center for Functional Materials, Wake Forest University, Winston-Salem, North Carolina 27109, United~States}
\author{Juan-Pablo~Correa-Baena}
\email{jpcorrea@gatech.edu}
\affiliation{School of Materials Science and Engineering, Georgia Institute of Technology, 771 Ferst Dr NW, Atlanta, GA~30332, United~States}
\author{Ajay~Ram~Srimath~Kandada}
\email{srimatar@wfu.edu}
\affiliation{Department of Physics \& Center for Functional Materials, Wake Forest University, Winston-Salem, North Carolina 27109, United~States}
\date{\today}

\begin{abstract}
The layered structure of two-dimensional metal halide perovskites (MHPs) consisting of an ionic metal halide octahedral layer electronically separated by an elongated organic cation, exhibits strong coupling between high-binding-energy excitons and low-energy lattice phonons. Photo-excitations in these systems are believed to be exciton-polarons–Coulombically bound electron-hole pairs dressed by lattice vibrations. Recent students reveal lattice-driven excitonic processes, highlighting the structural dependency of exciton-phonon coupling. Understanding and controlling the structural and chemical factors that govern this interaction is crucial for optimizing exciton recombination, transport, and many-body interactions. Our study examines the role of the organic cation in a prototypical 2D-MHP system,  phenylethylammonium lead iodide, \ce{(PEA)2PbI4}, and it’s halogenated derivatives, \ce{(F/Cl-PEA)2PbI4}. These substitutions allow us to probe polaronic effects while maintaining the average lattice and electronic structure. Using resonant impulsive stimulated Raman scattering (RISRS), we analyze the metal-halide sub-lattice motion coupled to excitons. We apply formalism based on a perturbative expansion of the nonlinear response function on the experimental data to estimate the Huang-Rhys parameter, $\mathbf{S}=\frac{1}{2}\Delta^2$, to quantify the lattice displacement ($\Delta$) due to exciton-phonon coupling. A direct correlation emerges between lattice displacement and octahedral distortion, with F-PEA experiencing the largest shift and Cl-PEA exhibiting the least, significantly influencing the fine structure features in absorption. Additionally, 2D electronic spectroscopy reveals that F-PEA, with the strongest polaronic coupling, exhibits the least thermal dephasing, supporting the polaronic protection hypothesis. Our findings suggest that systematic organic cation substitution serves as a tunable control for the fine structure in 2D-MHPs, and offers a pathway to mitigate many-body scattering effects by tailoring the polaronic coupling.
\end{abstract}

\maketitle
\section{Introduction}
Engineering the coupling between electronic excitations and lattice vibrations is crucial for optimizing the electronic, optical, and thermal properties of materials. This coupling significantly influences key optoelectronic parameters, including carrier mobility~\cite{wehrenfennig2013high, yi2016intrinsic, raimondo2013exciton, gong2024boosting}, thermal conductivity~\cite{tong2019comprehensive, ordonez2011effect, lin2019ultralow, acharyya2020intrinsically}, optical absorption coefficients~\cite{mishra2018exciton, shree2018observation, nguepnang2021electron, sarma1985optical}, luminescence quantum yields~\cite{huang2023origin, luo2021regulating, zhao2023pressure}, and exciton binding energies~\cite{katan2019quantum, mishra2018exciton, wang2017influence, marjit2023impacts}. Consequently, understanding and controlling the structural and chemical factors that govern electron-phonon coupling is essential for advancing the efficiency of optoelectronic devices. Furthermore, the interplay between charge carriers and lattice fluctuations can give rise to emergent phenomena relevant to quantum technologies, while simultaneously disrupting quantum coherence by opening multiple scattering pathways~\cite{kaer2014decoherence, quiros2024strong, song2024quantum, fu2023carriers, roche2005quantum, sun2021phonon, nakamura2024quantum}. Therefore, tailoring electron-phonon coupling is imperative to achieve the desired material functionality for specific applications.

In semiconducting materials, polaronic effects are among the most relevant and striking manifestations of electron-lattice coupling~\cite{li2020strong, buizza2021polarons, shi1997polaron, franchini2021polarons}. These effects are particularly pronounced in polar semiconductors, where the interaction with lattice vibrations fundamentally shapes their electronic and optical properties~\cite{mahanti1972effective, matsuura1980optical}. When a charge carrier moves through an ionic or polar crystal, its electric field interacts with the lattice atoms or ions, displacing them from equilibrium positions. This creates a localized distortion, or polarization, around the carrier, effectively coupling the charge carrier to the lattice vibrations~\cite{lanzani2007coherent, emin2013polarons}. This combined entity of the charge carrier and its associated lattice distortion is known as a polaron. Polaronic coupling is typically quantified through a dimensionless coupling parameter, $\alpha$, estimated using the known values of the phonon energy and the dielectric response function~\cite{emin1993optical}. A more direct estimation can be made by measuring the carrier's effective mass, where the mass enhancement factor directly reflects the strength of electron-phonon coupling~\cite{petrov2020ruthenium, pipa2001electron, tiras2013effective}. Alternatively, optical signatures such as vibronic replicas in absorption spectra~\cite{manrho2022optical, zaitsev2000vibronic, dzhagan2018vibrational} or characteristic lineshapes in vibrational spectra~\cite{luer2009coherent, lanzani2007coherent} can be used to quantitatively determine the degree and nature of lattice distortion induced by the electronic excitation.

In nanostructured material systems, electronic dynamics are often governed by Coulomb interactions between electrons and holes, forming bound pairs known as excitons~\cite{wheeler2013exciton, scholes2006excitons, burgos2020exciton}. Consequently, exciton-phonon coupling becomes a more relevant consideration. Although excitons are charge-neutral, the inherent and distinct lattice coupling of the constituent electrons and holes can still lead to substantial exciton-phonon interactions. This is particularly significant for materials such as III-V semiconductors hosting excitons with large Bohr radii (small exciton binding energies) and the exciton couples with long-range optical phonons~\cite{miyata2017large, bao2023evidence, puppin2020evidence}. In contrast, for localized excitons with small radii, such as those in organic semiconductors, the coupling of excitons with high-frequency molecular vibrations is primarily driven by local atomic reorganization, quantified by the Huang-Rhys factors~\cite{ghosh2020excitons, spano2010spectral, alexandrov2008polarons}. 

An intriguingly large coupling between high-binding energy excitons and low-energy lattice phonons is observed in some emerging organic-inorganic hybrid materials~\cite{yu2022excitons, powers2024coherent, ni2017real}. Prominent examples include hybrid 2D metal halides~\cite{tao2022dynamic, mauck2019excitons}, which are derivatives of hybrid metal halide perovskites. In these systems, electronic excitations are confined within a quasi-2D layer of metal halide octahedra separated by long organic cations~\cite{diab2016narrow, gauthron2010optical, even2014understanding}. We had previously suggested, based on the analysis of temperature-dependent optical lineshapes, that such intriguing behavior possibly originates from an interplay of long-range and short-range exciton-lattice couplings mediated by the organic-inorganic interactions unique to these material systems~\cite{srimath2020exciton}, but is yet to be fully substantiated. 

The mechanism of exciton-lattice coupling and the manifestation of polaronic effects in this class of materials has been extensively studied over the past decade~\cite{biswas2024exciton, srimath2020exciton, thouin2019phonon, fu2021electronic, duan20242d}, yet a consensus remains elusive. In particular, the role of the organic cation in influencing complex polar lattice dynamics and the lattice dressing of electronic excitations is not well understood. Organic cations, which occupy the vacancy between the octehedra, are typically terminated with an amine group and interact with halogen atoms through hydrogen bonding~\cite{zhouspacer, ou2025role}, which makes it imperative to consider how their chemical and structural properties impose local strain and lattice distortions. These interactions affect lattice parameters, especially the octahedral angles~\cite{ghosh2018mixed, liu2022effects, wu2020study}, which vary with the choice of organic cation. Additionally, the localized nature of these interactions introduces measurable variations in lattice parameters, contributing to static lattice disorder. While the impact of such lattice deviations on the electronic structure~\cite{giorgi2014cation, ghosh2018mixed, maheshwari2018computational, maheshwari2019effect}, bandgap~\cite{zibouche2020structure, motta2015revealing, mladenovic2018effects, marchenko2021relationships, fraccarollo2016ab}, and exciton binding energies~\cite{passarelli2020tunable, selig2017organic, motta2016effects, chen2015rotational} has been explored through both computational and experimental approaches, their effects on phonon energies, exciton-phonon interactions, and polaronic couplings remain underexplored. Furthermore, the substitution of organic cations simultaneously modifies both phonon couplings and intrinsic electronic properties~\cite{straus2018electrons, ruan2023impact, herz2018lattice, gallop2018rotational, munson2021influence}, complicating the precise determination of mechanisms governing exciton-lattice coupling. This complexity underscores the importance of preserving the integrity of excitons in such studies.

We focus our studies on a prototypical 2D-MHP system, phenylethylammonium lead iodide, \ce{(PEA)2PbI4}, which has been extensively studied in the field, and its halogenated derivatives, fluorinated/chlorinated phenylethylammonium lead iodide, \ce{(F/Cl-PEA)2PbI4}. \ce{(PEA)2PbI4} provides us with an ideal platform to study exciton-phonon interactions, due to its unique exciton-lattice couplings and complicated excitonic fine structure at low temperature, composed of four distinguishable peaks with equal spacing of about $\Delta \approx 35-40$ meV \cite{srimath2020exciton}. It has been shown through resonant impulsive stimulated Raman scattering (RISRS) that different excitons induce unique lattice reorganization, which is proof of polaronic character \cite{thouin2019phonon}, suggesting that the spectral fine structure portrays a family of coexisting excitons, with binding energy offset by $\Delta$, each with a distinct lattice dressing \cite{srimath2020exciton, thouin2019enhanced, thouin2018stable, neutzner2018exciton, thouin2019phonon, thouin2019electron}. Crystallography and linear spectroscopy provide sufficient evidence that the excitonic landscape is remaining largely intact across the three variations, allowing us to only comment on changes concerning the organic-inorganic framework as a whole. Utilizing RISRS, we establish that the organic cation influences the nature of coupling between the exciton and local lattice vibrations, directly impacting the polaronic character of the excitons. Our findings indicate that a systematic substitution of the organic cation could provide researchers a control knob to tune the complicated fine structure present and mitigate many-body scattering effects in these materials by altering the degree of polaronic coupling.
\section{Results}

\begin{figure}[h] 
    \centering
    \includegraphics[height=8cm]{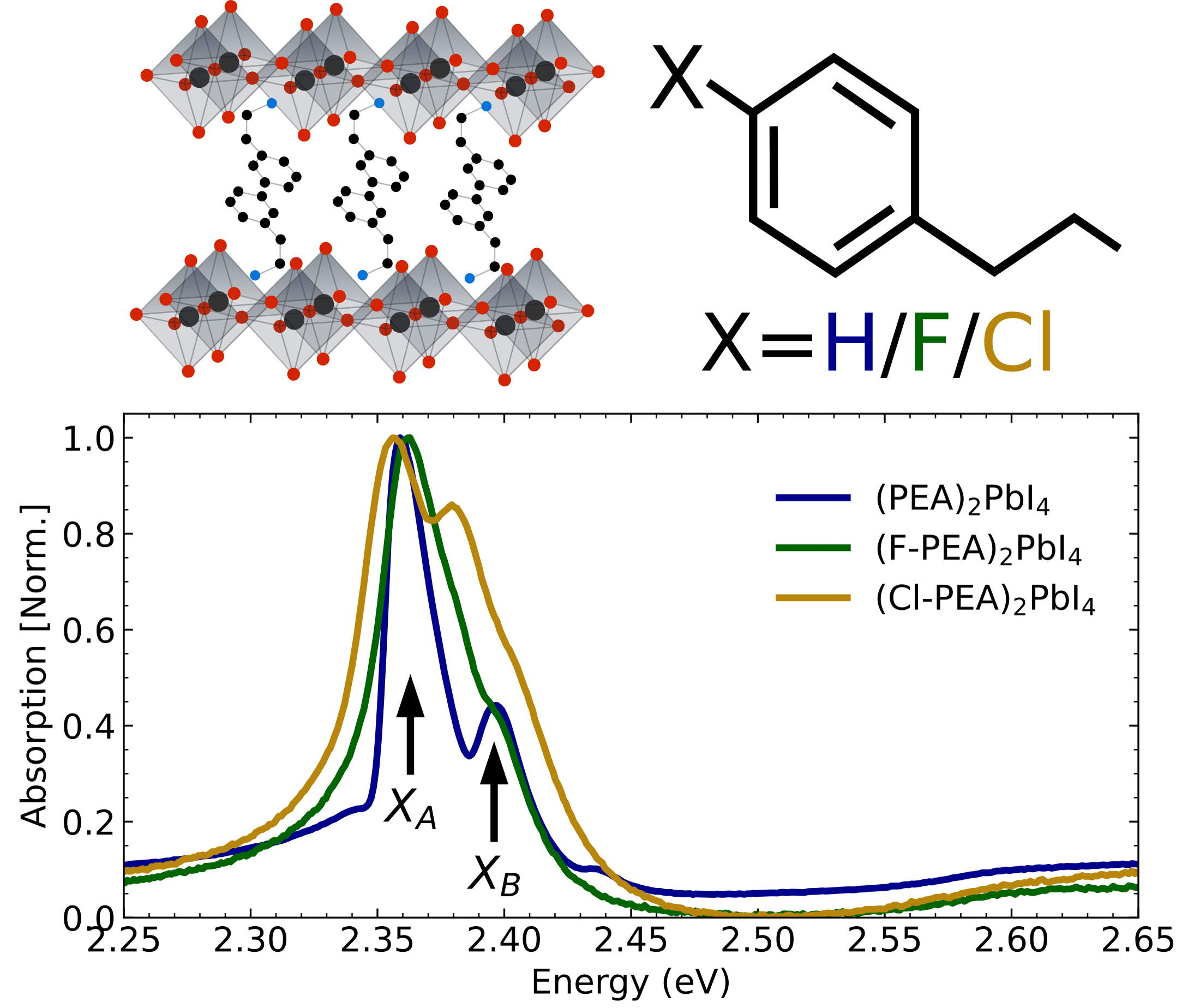}
    \caption{(a)-(c) Structure schematic of \ce{(PEA)2PbI4}, \ce{(F-PEA)2PbI4}, and \ce{(Cl-PEA)2PbI4} respectively. (d) Linear absorption spectra at 15 K for all three materials.}
    \label{fig:abs}
\end{figure}

It is well established that the choice of organic cation influences both the lattice parameters and electronic structure~\cite{maheshwari2018computational, giorgi2014cation, marchenko2021relationships}. In this study, we selected three cations—unsubstituted phenethylammonium (4-H), fluorine-substituted phenethylammonium at the fourth position (4-F), and chlorine-substituted phenethylammonium at the fourth position (4-Cl)—to minimize disruption to the lead-halide octahedral framework (Figs.~\ref{fig:abs}(a)–(c)). This approach ensures that the electronic structure and excitonic characteristics remain largely unaffected. Indeed, as shown in Fig.~\ref{fig:abs}(d), the linear absorption spectra of all samples exhibit qualitatively similar features, with the primary exciton peak and carrier continuum edge appearing at the same energy position. However, we observe discernible differences in the exciton fine structure among the samples, which we will explore later in this manuscript. Previous studies have suggested that such optical lineshape variations may originate from lattice interactions. While the organic cation substitution does not significantly alter the average crystal structure, it is expected to influence lattice flexibility, potentially modulating its dynamic properties.

Organic-inorganic interactions in perovskites and their derivatives are primarily governed by hydrogen bonding between the amine group of the organic cation and the halide ion in the inorganic framework~\cite{el2016hydrogen, el2016hydrogen, egger2016hybrid, lee2021dynamic}. These local interactions introduce static disorder by altering octahedral parameters, leading to average distortions in the unit cell. Such distortions significantly influence the electronic structure~\cite{peksa2024engineering, knutson2005tuning, straus2022photophysics, dyksik2020broad, baranowski2019phase, zhong2024evolution}. X-ray diffraction allows us to assess the structural variations induced by the organic cation substitution. 

\begin{figure}[h] 
    \centering
    \includegraphics[height=10cm]{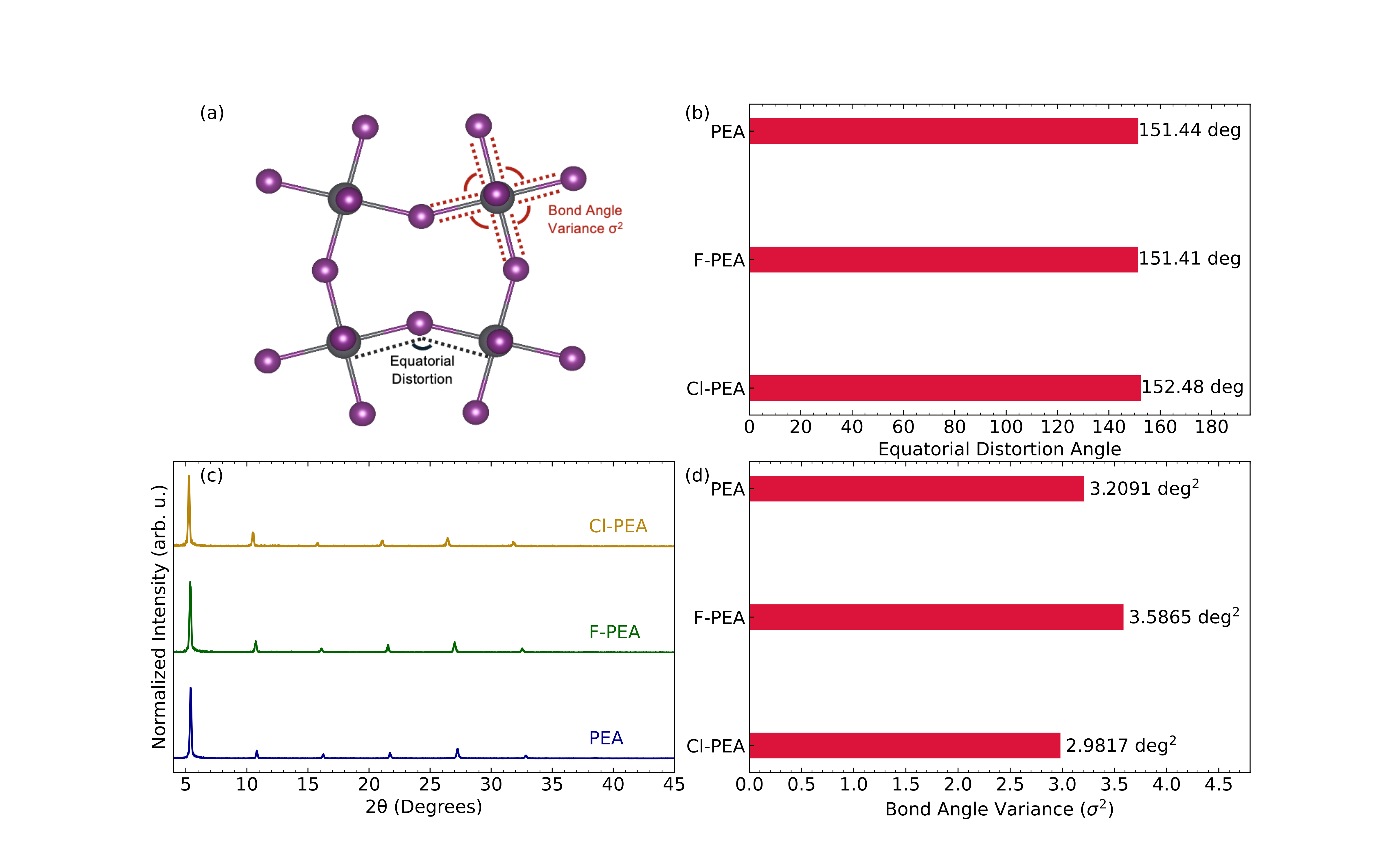}
    \caption{(a) Pictorial depiction of the equatorial distortion angle and the bond angle variance ($\sigma^2$) as a measure of deviation from the ideal bond angle of 90\degree. Graphical representations of (b) the equatorial distortion angle, (c) the XRD patterns, and (d) the bond angle variance ($\sigma^2$) are shown for all three cations.}
    \label{fig:crystal}
\end{figure}

The X-ray diffraction patterns measured from the 2D MHP thin films used in this work are shown in Fig.~\ref{fig:crystal}(c). These materials exhibit a preferred out-of-plane orientation, as indicated by the characteristic Bragg peaks corresponding to the \{002\} family of planes~\cite{steele2023giwaxs, wright2022influence}. All the materials form oriented Ruddlesden-Popper phases with varying interplanar distances, as evident from the spacing between the \{002\} planes. Among the samples, \ce{(Cl-PEA)2PbI4} exhibits the greatest interplanar distance, while \ce{(PEA)2PbI4} has the smallest. The diffraction patterns obtained for the thin films correlate well with previously reported single-crystal X-ray diffraction~\cite{straus2019longer}, as shown by the comparison between Fig.~\ref{fig:crystal}(c) and Figure S1. The calculated thin-film XRD closely aligns with the simulated diffractogram for the $<$001$>$ facet orientation, obtained from the single-crystalline data available in the literature~\cite{straus2019longer}. 

To explore the structural differences induced using different cations, we focus on two parameters that are relevant for the electronic properties: equatorial distortion and bond angle variance, illustrated schematically in Fig.~\ref{fig:crystal}(a). We estimate these parameters from the single-crystal data, which was shown to accurately represent the average structure of our materials.  The electronic structure and bandgap are particularly sensitive to equatorial distortion~\cite{marchenko2021relationships}, which, as shown in Fig.~\ref{fig:crystal}(b), remains largely consistent across the three samples studied. This aligns with the linear absorption data in Fig.~\ref{fig:abs} that suggests similar exciton and bandgap energies across the samples.

To quantify the angular distortions within the octahedra, we compute the Bond Angle Variance ($\sigma^2$), shown in Fig.~\ref{fig:crystal}(d), as a measure of deviation from the ideal bond angle of 90\degree. This metric has been widely used to assess octahedral distortions in low-dimensional perovskites, providing a quantitative means to correlate structural variations with differences in optical properties~\cite{yan2022electronic, fu2020cation}. The bond angle variance ($\sigma^2$) is graphically represented in Fig.~\ref{fig:crystal}(a) and is calculated using the following formula:

\begin{gather}
      \sigma^2 =\frac{1}{m-1} \sum_{i=1}^{m} (\phi_i - \phi_0)^2
\label{eq:A}
\end{gather}

Where $\phi_i$ is the ith bond angle and $\phi_0$ is the ideal bond angle for a regular octahedra (90\degree), which are estimated from the crystallographic data. Fig.~\ref{fig:crystal}(d) shows the bond angle variance estimated for the three samples. 
The F-substituted cation \ce{(F-PEA)2PbI4} exhibits the greatest octahedral distortion, followed by \ce{(PEA)2PbI4} and then \ce{(Cl-PEA)2PbI4}, which shows the least distortion. The effect of octahedral distortion in 2D perovskites is not well understood, but it is believed that the disordered crystal structure has a role in the coupling enhancement between excitations and lattice distortions \cite{shao2022unlocking, cherrette2023octahedral, halder2017octahedral}. We will comment on the bond angle variance differences later in the discussion section of this manuscript. 

To investigate the impact of cation substitution on the motion of the metal-halide sublattice—specifically, the vibrational modes that couple directly with the exciton—we performed resonant impulsive stimulated Raman scattering (RISRS) measurements on \ce{(PEA)2PbI4} and its halogenated derivatives, \ce{(F-PEA)2PbI4} and \ce{(Cl-PEA)2PbI4}. In this technique, an ultrashort optical pulse impulsively excites the lattice, generating a coherent vibrational wavepacket—a collective superposition of all Raman-active vibrational modes. This wavepacket then oscillates along the potential energy surfaces of both the ground and excited electronic states, following the vibrational coordinates defined by the coupled phonon modes.

\begin{figure}[h]
    \centering
    \includegraphics[height=8cm]{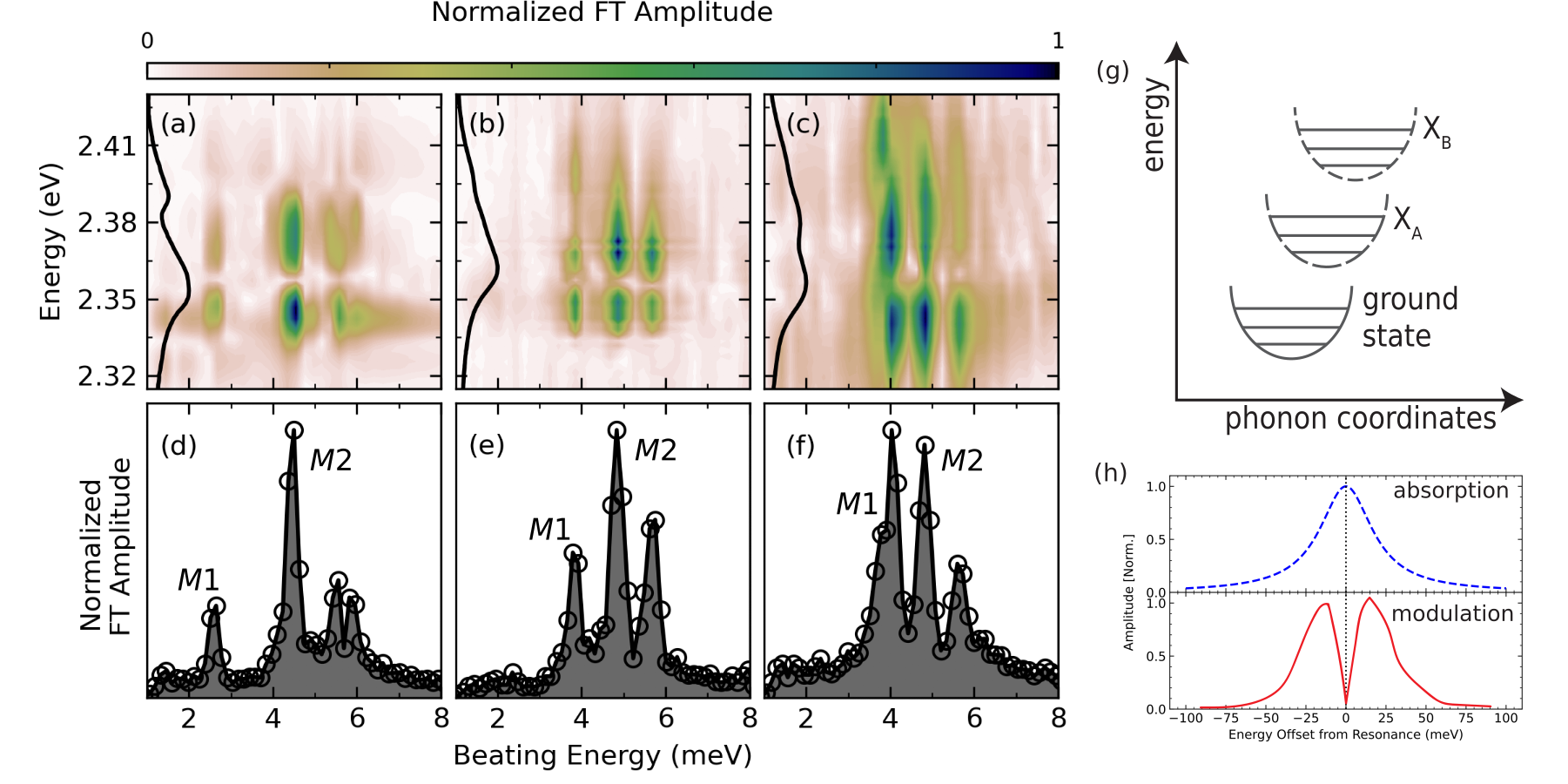}
    \caption{(a)-(c) Linear absorption (black line) and beating spectra as a function of detection energies for (a) \ce{(PEA)2PbI4}, (b) \ce{(F-PEA)2PbI4}, and (c) \ce{(Cl-PEA)2PbI4}. (d)-(f) Probe-energy-integrated vibrational spectra for \ce{(PEA)2PbI4}, \ce{(F-PEA)2PbI4}, \ce{(Cl-PEA)2PbI4} respectively. All measurements were taken at 15 K. (g) Schematic of potential energy surfaces, where the potential energy surfaces of $X_A$ and $X_B$ are composed of distinct vibrational manifolds. (h) Schematic of the amplitude spectrum of phonon coherences induced by resonant impulsive stimulated Raman scattering (RISRS) at the excitonic resonance. }
    \label{fig:beating}
\end{figure}

The resulting modulation of the complex refractive index manifests as a time-dependent oscillatory signal in the differential transmission spectrum (Fig. S2). By performing a Fourier transform along the time axis, we extract the RISRS modulation maps, shown in Figs.~\ref{fig:beating}(a)-(c). It is important to note that the transmission modulations induced by the coherent vibrational wavepacket appear exclusively at specific probe wavelengths, exhibiting a distinct lineshape. The nature and implications of these lineshapes will be discussed in detail later in this manuscript.

Focusing exclusively on \ce{(PEA)2PbI4}, we can draw key insights from the beating map displayed in Fig.~\ref{fig:beating}(a). In this figure, the black line represents the absorption spectrum, providing a reference for the excitonic resonances. Notably, the lower-energy exciton ($X_A$) exhibits a more pronounced modulation due to phonon interactions compared to the higher-energy exciton ($X_B$). This is evident from the characteristic dip in the modulation amplitude observed at the energy of $X_A$, signaling stronger exciton-phonon coupling at this resonance.

To further analyze the phonon contributions, we integrate the data over the energy axis, yielding the Raman spectra shown in Fig.~\ref{fig:beating}(d). This spectrum is analogous to a continuous wave resonant Raman spectrum albeit obtained via impulsive excitation. The RISRS spectrum reveals the presence of at least four distinct phonon modes. The data for \ce{(PEA)2PbI4} obtained in this study aligns well with previously reported Raman spectra from Thouin et al. \cite{thouin2019phonon}. In that work, we demonstrated that different excitons drive distinct lattice reorganizations, thereby providing direct experimental evidence of their polaronic nature. That study, supported by density functional theory (DFT) calculations, established that all identified phonon modes in \ce{(PEA)2PbI4} and below 10\,meV originate from vibrations within the lead-iodide octahedral network.

We observe phonon modes within a similar energy range across all samples, as evident in the integrated RISRS spectra shown in Fig.~\ref{fig:beating}(e) and (f). Given the energy range of these spectral modes, we attribute the observed phonons to the motion of the \ce{PbI4}$^{2-}$ framework. Interestingly, although the inorganic sub-lattice remains largely intact across the different derivatives, distinct variations in the RISRS response are apparent. A particularly notable trend across both \ce{(PEA)2PbI4} derivatives is the blue shift of the lowest-energy phonon, accompanied by an increase in its intensity as we move from the \ce{(F-PEA)2PbI4} compound to \ce{(Cl-PEA)2PbI4}. At present, crystallographic analysis does not reveal the structural variable responsible for this shift, leaving its origin an open question. For the sake of further discussion in this manuscript, we label the dominant modes as $M1$ and $M2$, which have different energies in each material, tabulated in Tab.~\ref{tab:modes}. 

\begin{table}[]
    \centering
    \begin{tabular}{c|c|c|c}
         Mode & \ce{PEA} & \ce{F-PEA} & \ce{Cl-PEA} \\
         \hline $M1$ & 2.6meV & 3.8 meV & 4 meV \\
         $M2$ & 4.4meV & 4.9 meV & 4.9 meV  
    \end{tabular}
    \caption{Dominant phonon mode ($M1$ \& $M2$) energies for \ce{(PEA)2PbI4}, \ce{(F-PEA)2PbI4}, and \ce{(Cl-PEA)2PbI4}.}
    \label{tab:modes}
\end{table}

Beyond changes in the eigenmode energies, we now examine how the RISRS response evolves across different probe energies. Vertical cuts along the beating energy axis of the probe-energy-resolved beating maps in Fig.~\ref{fig:beating}(a)–(c) exhibit a characteristic lineshape, schematically illustrated in Fig.~\ref{fig:beating}(h). Notably, at the exciton absorption peak energy, a distinct dip appears in the modulation spectrum. The observed dual-peak lineshape is a hallmark of vibrational wavepacket dynamics, where the exciton energy is modulated by the motion of a vibrational wavepacket along a real-space vibrational coordinate associated with the coherently excited phonon~\cite{luer2009coherent, lanzani2007coherent}. This characteristic spectral feature is typically seen in localized electronic excitations of organic chromophores coupled to molecular vibrations~\cite{arpin2015spectroscopic, jumper2016broad, johnson2014photocycle, polli2010conical, kahan2007following, vos1993visualization, pollard1992theory, mcclure2014coherent}. Similar spectral responses have been reported in 2D metal halides like \ce{(PEA)2PbI4} and its derivatives, which have been interpreted—both by us~\cite{thouin2019phonon} and others~\cite{biswas2024exciton, fu2021electronic}—as signatures of polaronic coupling between excitons and lattice phonons. In this framework, the excitonic potential energy surface (PES) is displaced along the vibrational coordinates defined by the lattice phonons. The modulation strength and lineshape symmetry are linked to the extent of this displacement in the excited-state PES and the degree of anharmonicity in the potential.

\begin{figure}[h]
    \centering
    \includegraphics[height=7cm]{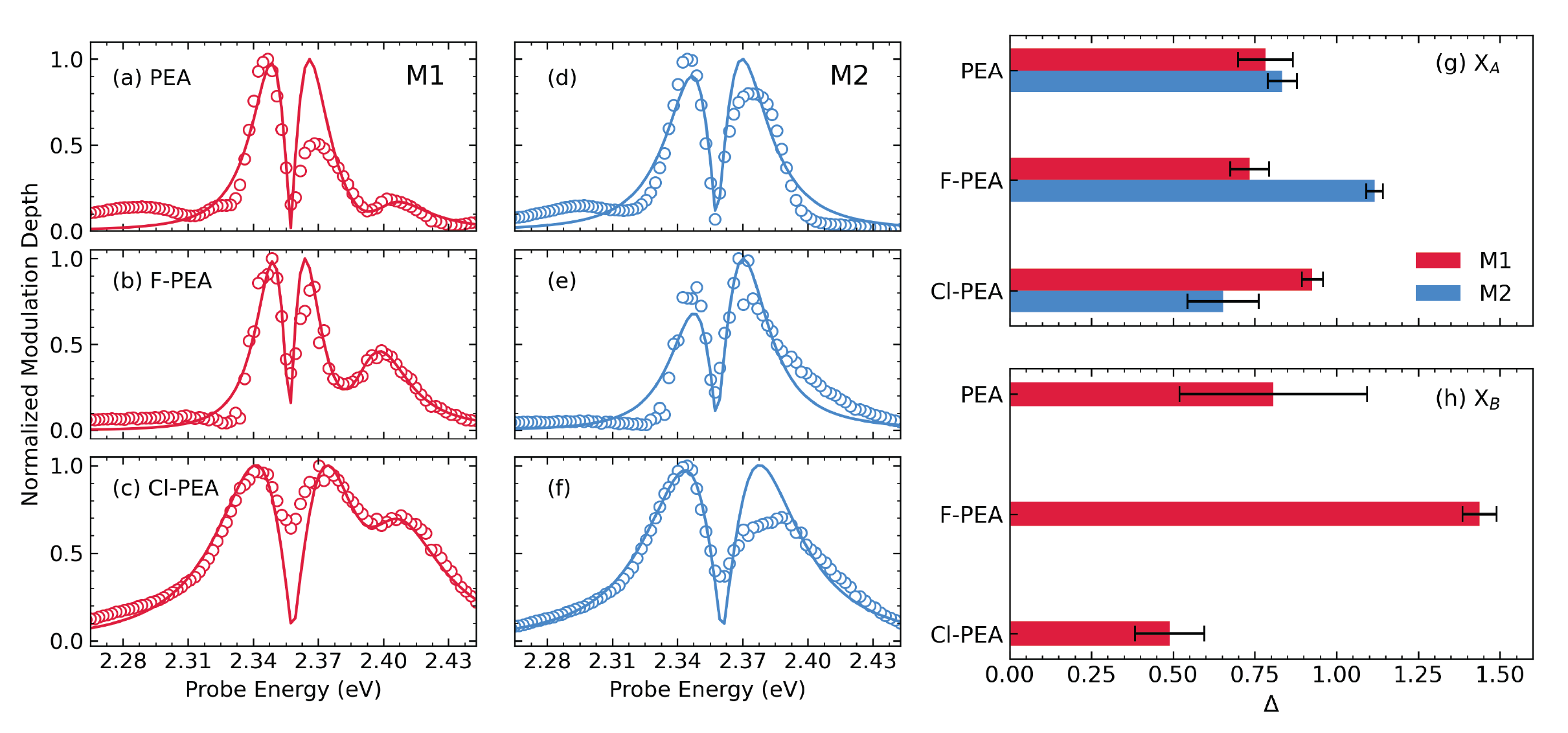}
    \caption{Spectral cuts of RISRS beating map and fitting results using Equations \ref{eq:M} and \ref{eq:A}. This figure shows the modulation depth for phonon mode 1 ($M1$) and phonon mode 2 ($M2$), where spectral cuts were taken at (a) 2.3-2.9 meV and (d) 4.1-4.8 meV for \ce{(PEA)2PbI4}; (b) 3.6-4.1 meV and (e) 4.5-5.3 meV for \ce{(F-PEA)2PbI4}; (c) 3.6-4.4 meV and (f) 4.6-5.2 meV \ce{(Cl-PEA)2PbI4}. The data is represented with open circles and the fit is the solid line. The offset between the ground and excited state potential energy surfaces calculated from the Huang-Rhys fitting parameter for (g) exciton A ($X_A$) and (h) exciton B ($X_B$).}
    \label{fig:FCS_fits}
\end{figure}

The RISRS maps in Figs.~\ref{fig:beating}(a)–(c) reveal that the modulation associated with $M2$ occurs exclusively at the energy of the dominant peak, $X_A$, in the linear absorption of all samples. In contrast, the modulation due to $M1$ appears not only at $X_A$ but also at the higher-energy absorption peak, labeled $X_B$. This observation suggests the presence of two distinct excitonic states, $X_A$ and $X_B$, each interacting uniquely with the lattice phonons $M1$ and $M2$, and with distinct lattice displacements (Fig.~\ref{fig:beating}(g)). To further quantify these differences across the samples, we analyze the modulation amplitude spectra for $M1$ and $M2$, by taking vertical cuts of the RISRS map at the respective energies (see Fig.~\ref{fig:FCS_fits}(a)-(f) \& Fig. S3).

Researchers have measured similar amplitude profiles due to vibrational wavepackets since the 1990s, but until recently qualitative information has been difficult to extract \cite{pollard1992analysis, mukamel1995principles}. Turner et. al. \cite{barclay2022characterizing, turner2020basis, arpin2021signatures},
developed an analytical model for femtosecond coherence spectra (FCS), which they deﬁne as the Fourier-domain amplitude and phase proﬁles, A($\omega$) and $\phi$($\omega$), respectively, as a function of detection frequency, $\omega$, for a chosen oscillation frequency, $\omega_0$. The model begins with the displaced harmonic oscillator model, where both the ground and excited electronic states are harmonic potential energy surfaces of frequency $\omega_0$, offset horizontally by a displacement $\Delta$, which is directly related to the Huang-Rhys factor by $\mathbf{S}=\frac{1}{2}\Delta^2$. The vertical offset is represented by the exciton resonance energy $\omega_{eg}$. A doorway-window approach is used, where the 
vibrational wavepacket oscillating within the excited state PES is projected onto the ground state PES, and the Frank-Condon wavefunction overlap is determined~\cite{mukamel1995principles, turner2020basis}.

The starting point of the model is the transient absorption data, $S(\omega,\tau_2)$, where $\tau_2$ is the pump probe delay and $\omega$ is the probe (detection) energy axis. A Fourier transform over the time interval results in a complex valued spectrum, $M(\omega,\omega_2)=\mathcal{F}[S(\omega,\tau_2)]$. The amplitude and phase profiles of a specific mode, with frequency $\omega_0$, are functions of detection frequency ($\omega$), and given by, $A(\omega)=M(\omega,\omega_2)|_{\omega_2=\omega_0}$. The two relevant equations for this model are: 

\begin{equation}
    M(\omega ; \omega_0) = e^{-2\mathbf{S}}\sum_{n,m}^{\infty} m! \mathbf{S}^{2n+m+1}A_{n,m}(\mathbf{S})A_{n+1,m}(\mathbf{S})\\
\left[\frac{1}{\omega-\omega_{n+1,m}+i\gamma/2}-\frac{1}{\omega-\omega_{n,m}-i\gamma/2}  \right]
\label{eq:M}
\end{equation} 

where the auxiliary functions are given by:

\begin{gather}
      A_{a,b}(\mathbf{S}) =\sum_{j=0}^{min[a,b]} \frac{(-1)^j \mathbf{S}^{-j}}{j!(a-j)!(b-j)!}   
\label{eq:A}
\end{gather}

\begin{figure}[h]
    \centering
    \includegraphics[width=11cm]{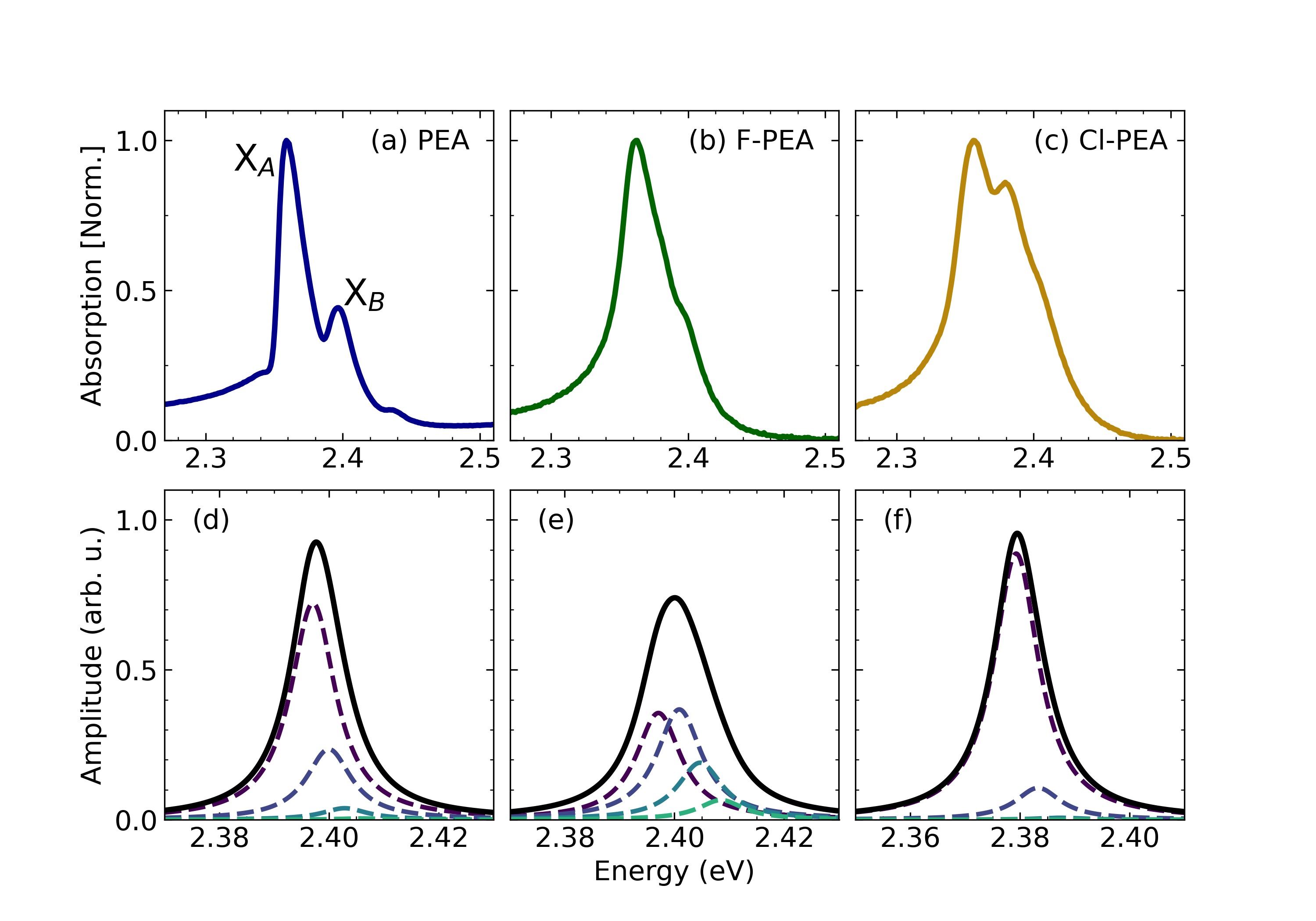}
    \caption{The offset between the ground and excited state potential energy surfaces calculated from the Huang-Rhys fitting parameter for (a) exciton A and (b) exciton B. Linear absorption spectrum of (c) \ce{(PEA)PbI4}, (d) \ce{(F-PEA)PbI4}, and (e) \ce{(Cl-PEA)PbI4}. (f)-(h) Represent the collection of Lorentzian peaks, weighted by $F_{n0}=\frac{e^{-\mathbf{S}}\mathbf{S}^n}{n!}$ where n denotes the vibrational transition, that make up the $X_B$ resonance, and the sum of those peaks (black line).}
    \label{fig:Huang-Abs_Lorentz}
\end{figure}

In these expressions, $m$ and $n$ index the vibrational eigenstates of the ground and excited electronic states respectively, and $\gamma$ is the dephasing parameter. Physically Equations \ref{eq:M} and \ref{eq:A} represent that each FCS spectra is a collection of Lorentzian peaks, where each peak is broadened by $\gamma$ and weighted by an auxiliary function, see Fig.~\ref{fig:Huang-Abs_Lorentz}. The peaks overlap and lead to lineshapes similar to Fig.~\ref{fig:beating}(h). The fit parameters for this model are \{$\mathbf{S}$, $\gamma$, $\omega_{eg}$\} where the phonon frequency, $\omega_0$, is defined directly from the measured spectra. Value restrictions can be placed on both $\gamma$ and $\omega_{eg}$ because they can be approximated from the linear absorption linewidths and the energy of the exciton resonance, respectively, leaving the only true fitting parameter to be $\mathbf{S}$. 

This model was able to reproduce the $M2$ modulation spectra of all three compounds shown in Fig.~\ref{fig:FCS_fits}(d)-(f), which follow the expected dual peak lineshape. However, as can be seen in Fig.~\ref{fig:FCS_fits} the lowest energy phonon mode (red) deviates from the two-peak structure seen in the second mode (blue), and a third peak is emerging at higher energies. For these low energy phonon modes, we expanded the model to include two distinct excited state potential energy surfaces, each for $X_A$ and $X_B$ respectively, see Equation \ref{eq:M_expanded} and Fig.~\ref{fig:beating}(g), where the relative weight of contribution for the $X_A$ and $X_B$ PESs were determined using the absorption spectrum (A = intensity of $X_A$ resonance and B = intensity of $X_B$ resonance). 

\begin{equation}
    M_{total}(\omega ; \omega_0) = A\cdot M_{X_A} + B\cdot M_{X_B}
\label{eq:M_expanded}
\end{equation} 

This analytical model enabled the extraction of the Huang-Rhys parameter, which is related to the equilibrium offset between the ground and excited state potential energy surfaces, and hence the extent of polaronic coupling. We highlight that the FCS model is the most accurate estimation of the Huang-Rhys parameter because it provides direct access to time-resolved exciton-phonon dynamics, unlike Raman or sideband methods which infer the Huang-Rhys parameter through steady-sate or frequency domain data. Additionally, the FCS method addresses the fact that phonon replicas in the Raman or photoluminescence spectra are difficult to deconvolute, as it resolves the exciton-phonon interactions temporally, enabling one to distinguish phonon modes with femtosecond precision, which is necessary for these systems as the phonon energies of interest are below 10 meV. In Figures ~\ref{fig:FCS_fits}(g) and (h), we show the Huang-Rhys parameter associated with the two dominant vibrational modes ($M1$ and $M2$) for two distinct excitonic resonances (labeled $X_A$ and $X_B$) in the optical spectrum. 

\section{Discussion}
For all systems considered, the PES of $X_A$ is displaced along coordinates associated with $M1$ and $M2$, while $X_B$ shifts exclusively along $M1$. We first look at the displacement $\Delta$ of $X_A$ along the $M2$ mode, which has been identified as the dominant mode in the photoexcitation dynamics of the $X_A$ exciton~\cite{thouin2019phonon, srimath2020exciton}. Among the compounds studies, \ce{(F-PEA)2PbI4} exhibits the greatest dipplacment along $M2$, while \ce{(Cl-PEA)2PbI4} shows the least, with the non-halogenated compound displaying an intermediate value. Interestingly, the displacement of $X_A$ along $M1$ coordinate, and qualitatively follows the opposite trend, albeit to a lesser magnitude. Meanwhile, for $X_B$, which is displaced solely along $M1$, a similar trend emerges, with \ce{(F-PEA)2PbI4} demonstrating the most significant phonon dressing.

A clear and striking correlation emerges between the trends in exciton-phonon coupling parameters shown in Fig.~\ref{fig:FCS_fits}(g) and the crystallographic parameters, particularly the bond angle variance (BAV) in Fig.~\ref{fig:crystal}(d). When bond angles are close to their ideal values, the structure is more rigid and less susceptible to external stress or deformation. Conversely, a higher BAV indicates increased structural flexibility~\cite{bauchy2011angular, baur2019floppiness, marchenko2021relationships}, allowing the material to accommodate distortions induced by external forces. The lattice dressing effect in the presence of photocarriers, previously discussed as the polaronic effect, operates through Coulomb-mediated lattice deformations, which are directly linked to lattice flexibility. Our experimental results suggest that substituting \ce{PEA} with \ce{F-PEA} enhances the flexibility of the \ce{PbI4}$^{2-}$ sub-lattice, influencing the lattice dressing of excitons. In contrast, substitution with the chlorinated compound leads to a more rigid sub-lattice. Notably, while the overall electronic structure remains largely unchanged across the samples, subtle local interactions are fine-tuned by organic cation substitution. These structural modifications have significant and observable effects on key spectral characteristics, which we will explore in the following sections.

The complex spectral fine structure, characterized by multiple transitions at the exciton energy, has been widely reported in numerous 2D metal-halide perovskites (2D-MHPs) incorporating various organic cations and halogens \cite{even2014understanding, rojas2023many, srimath2022homogeneous, ghosh2020polarons, franchini2021polarons}. However, the origin of this fine structure remains an open question, with ongoing debate as to whether the observed transitions correspond to distinct excitonic states or a single transition exhibiting vibronic progression.

In our previous work, we argued that each observed peak in absorption and photoluminescence (PL)—denoted here as $X_A$ and $X_B$ -- corresponds to distinct excitonic states. However, we also noted the possible presence of additional peaks, obscured within the broad linewidths of the linear response, which may be attributed to phonon replicas. Indeed, analysis based on the displaced oscillator model, used to interpret the Resonant Inelastic Scattering of Raman Signals (RISRS) response, predicts the presence of phonon replicas in the optical spectra. These replicas are separated by the characteristic phonon energies $M1$ and $M2$ and are weighted according to the estimated Huang-Rhys factors.

Notably, variations in the relative intensities and linewidths of the exciton fine structure can be observed across different samples (see Fig.~\ref{fig:abs} and Figs.~\ref{fig:Huang-Abs_Lorentz}(a)--(c)). Specifically, in \ce{(F-PEA)2PbI4}, the relative strength of $X_B$ appears quenched, whereas in \ce{(Cl-PEA)2PbI4}, it is enhanced. To further investigate how lattice displacement influences the oscillator strength of the $X_B$ resonance, we examine its underlying vibronic structure. Figures \ref{fig:Huang-Abs_Lorentz}(d)-(f) depict a collection of Lorentzian peaks weighted by the experimentally determined Franck-Condon factors, given by $F_{n0}=\frac{e^{-\mathbf{S}}\mathbf{S}^n}{n!}$, where $n$ denotes the vibrational transition that makes up the $X_B$ resonance. The black line represents the sum of these peaks. It is evident that an increase (decrease) in lattice displacement enhances (suppresses) the Franck-Condon overlap of higher-lying vibrational states, leading to a quenched (enhanced) transition. Organic cation substitution can thus offer effective ways of fine-tuning the spectral characteristics of the excitons in 2D-MHPs, which we have recently argues plays a critical role in enabling their application in polariton lasers~\cite{gomez2024multiple, quiros2024strong}.

Our findings further reinforce the hypothesis that photoexcitations in these materials must be understood within the exciton-polaron framework \cite{srimath2020exciton}, which becomes particularly relevant when considering many-body scattering effects. Within this framework, the exciton resides within a self-induced deformation cloud, where both phonon-phonon and exciton-phonon interactions contribute to the thermal dephasing of excitons. This dephasing manifests as a temperature-dependent increase in the homogeneous linewidth, $\gamma$. 

Previously, we noted that linear optical response does not provide direct access to $\gamma$~\cite{srimath2022homogeneous}, necessitating the use of two-dimensional electronic spectroscopy (2DES)~\cite{thouin2019enhanced, turner2011invited} instead. In Ref.~\citenum{thouin2019enhanced}, we employed 2DES to study thermal dephasing in \ce{(PEA)2PbI4}, allowing us to estimate the corresponding exciton-phonon interaction parameter. Importantly, we distinguish between the exciton-phonon coupling parameters associated with lattice dressing and exciton scattering, as they stem from inherently different physical processes.  

In this work, we extend our study to \ce{(F-PEA)_2PbI_4}, applying 2DES to determine the temperature dependence of $\gamma$. By assuming that line broadening results from exciton scattering with a single thermally populated phonon mode, we fit $\gamma$ to Eq. 3 to extract the effective energy ($E_{LO}$) of the scattered phonon and the effective interaction parameter ($\alpha_{LO}$). The fitting result reveal an effective interaction parameter of $\alpha_{LO}=1.58$ meV for \ce{(F-PEA)_2PbI_4}, see Fig. S4, which is significantly lower than the previously reported value of $\alpha_{LO}=33$ meV for \ce{(PEA)_2PbI_4} \cite{thouin2019enhanced}.

Notably, when comparing both the exciton-phonon interaction and Huang-Rhys parameters of the two materials, we observe an inverse correlation: a higher Huang-Rhys parameter—indicative of stronger polaronic character—corresponds to a lower exciton scattering interaction parameter. This finding highlights the interplay between polaron formation and exciton-phonon scattering in 2D-MHPs. These observations suggest that within the exciton-polaron framework, the dominant mechanism governing thermal dephasing is exciton-phonon scattering, rather than longitudinal optical (LO) phonons scattering with other LO phonons. Notably, in \ce{(F-PEA)2PbI4}, the enhancement of polaronic character, coupled with a reduction in the interaction parameter, supports the previous hypothesis that lattice phonon dressing provides a protective effect for excitations \cite{srimath2020exciton}. A comparative analysis of the interaction and Huang-Rhys parameters for \ce{(PEA)2PbI4} and \ce{(F-PEA)2PbI4} suggests that stronger lattice dressing effectively shields the exciton, mitigating scattering effects. This polaronic protection mechanism presents an intriguing avenue for material scientists seeking to suppress many-body scattering effects, potentially leading to significant enhancements in the performance of electronic and optoelectronic devices.
\section{Conclusions}

In summary, we have presented a comprehensive analysis detailing how exciton-phonon interactions are influenced by the organic cation, in 2D-MHP systems, specifically \ce{(PEA)2PbI4}, \ce{(F-PEA)2PbI4}, and \ce{(Cl-PEA)2PbI4}. Linear spectroscopy and crystallography were utilized to confirm that the organic cation substitutions preserve the exciton landscape, while RISRS, crystallography, and the estimation of the Huang-Rhys parameter established that the organic cation and hence the degree of octahedral distortion influences the nature of coupling between the exciton and local lattice vibrations, directly impacting the polaronic nature of the excitons. It is evident from the analysis that careful engineering of the organic cation and the organic-inorganic interactions may offer a design route to control the degree of polaronic character of excitons, and possibly provide an avenue to control the fine structure and weaken the many-body scattering effects of 2D-MHPs, which is crucial for advancements in material science.
\section{Author Contributions}
KAK performed the RISRS measurements, analyzed the data, and wrote the manuscript under the supervision of ARSK. MGD fabricated the samples and performed the crystallography analysis under the supervision of JPCB. ERG and KAK performed 2D spectroscopy under the supervision of ARSK. KBU participated in the collection of the RISRS data, and AE contributed to the FCS analysis. The project was conceived and coordinated by ARSK.  


\section{Acknowledgements}

JPCB and MGD were primarily supported by the National Science Foundation, Science and Technology Center Program (IMOD), under grant no. DMR-2019444. ARSK acknowledges funding from the National Science Foundation CAREER grant (CHE-2338663), start-up funds from Wake Forest University, funding from the Center for Functional Materials at
Wake Forest University. Any opinions, findings, and conclusions or recommendations expressed in this material are those of the authors(s) and do not necessarily reflect the views of the National Science Foundation. The authors are grateful to Professor Carlos Silva for providing access to the 2D spectroscopy system, and to Professor Stephen Winter and Ramesh Dhakal for insightful discussions. 


\providecommand{\noopsort}[1]{}\providecommand{\singleletter}[1]{#1}%


\clearpage

\begin{center}
    \noindent\large{\textbf{Supplemental Material: Fine-Tuning Exciton Polaron Characteristics via Lattice Engineering in 2D Hybrid Perovskites}} \\
  \vspace{0.5cm} 

  \normalsize
\end{center}

\subsection{Sample Preparation}
Glass slides were cleaned using sequential ultrasonic baths of acetone and isopropanol (IPA) for 15 minutes each, followed by nitrogen drying and UV-ozone treatment for 15 minutes. The perovskite precursor solutions were prepared by dissolving equimolar amounts of \ce{PbI2} (purity $>$99.99\%) and the corresponding organic cation (phenethylammonium iodide (purity $>$99.99\%), 4F-phenethylammonium iodide (purity $>$99.99\%), 4Cl-phenethylammonium iodide (purity $>$99.99\%))  in N,N-dimethylformamide (purity $>$99.98\%) at a concentration of 0.13 M. After stirring overnight, the perovskite films were deposited by dispensing 80 $\mu$L of the precursor solution onto the 2.54 cm$^2$ clean DBR, then spin-coated at 6000 RPM for 30s with an acceleration of 6000 RPM/s. Immediately after deposition, the films were thermally annealed at 100\degree C for 10 minutes.

 \subsection{X-Ray Diffraction (XRD)}

The calculated thin-film XRD presented in Fig. 2(c) closely aligns with the simulated diffractogram for the $<$001$>$ facet orientation (Fig.~\ref{fig:sim_XRD}), obtained from the single-crystalline data available in the literature~\cite{straus2019longer}. 

\setcounter{figure}{0}
\renewcommand{\figurename}{FIG.}
\renewcommand{\thefigure}{S\arabic{figure}}

 \begin{figure}[H] 
    \centering
    \includegraphics[width=9.5cm]{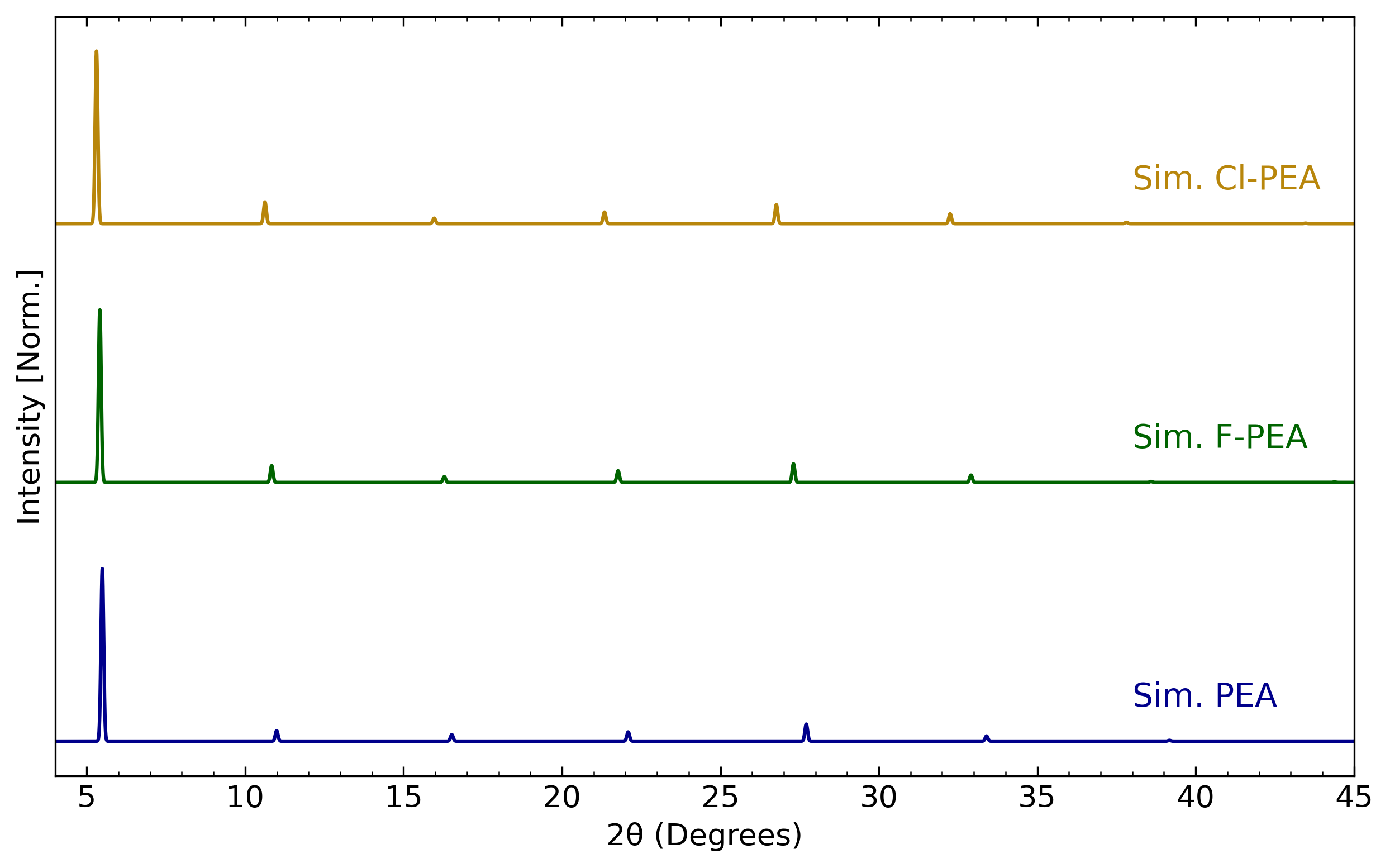}
    \caption{Simulated diffractogram for the $<$001$>$ facet orientation, obtained from Ref.~\citenum{straus2019longer}}
    \label{fig:sim_XRD}
\end{figure}

\subsection{Ultrafast Transient Absorption Spectroscopy}
Transient absorption spectroscopy measurements were performed using a pulsed femtosecond laser (Pharos Model PH1-20-0200-02-12, Light Conversion) emitting 1030nm pulses at 1kHz with a pulse duration of $\sim$200fs. The 2.88 eV pump beam was generated by feeding half of the laser output to a commercial optical parametric amplifier (Orpheus, Light Conversion) while 5 mW was focused onto a sapphire crystal to obtain a single filament white-light continuum covering the spectral range 450nm - 800nm for the probe beam. The probe beam transmitted through the sample was detected by a high speed camera (FLC 
3030, EB Stressing) in combination with a high resolution spectrometer (SpectraPro SP-2150, Princeton Instruments). All measurements were carried out in a vibration-free closed-cycle cryostation (Montana Instruments). 
\subsection{Resonant Impulsive Stimulated Raman Scattering}
\begin{figure}[H] 
    \centering
    \includegraphics[width=9.5cm]{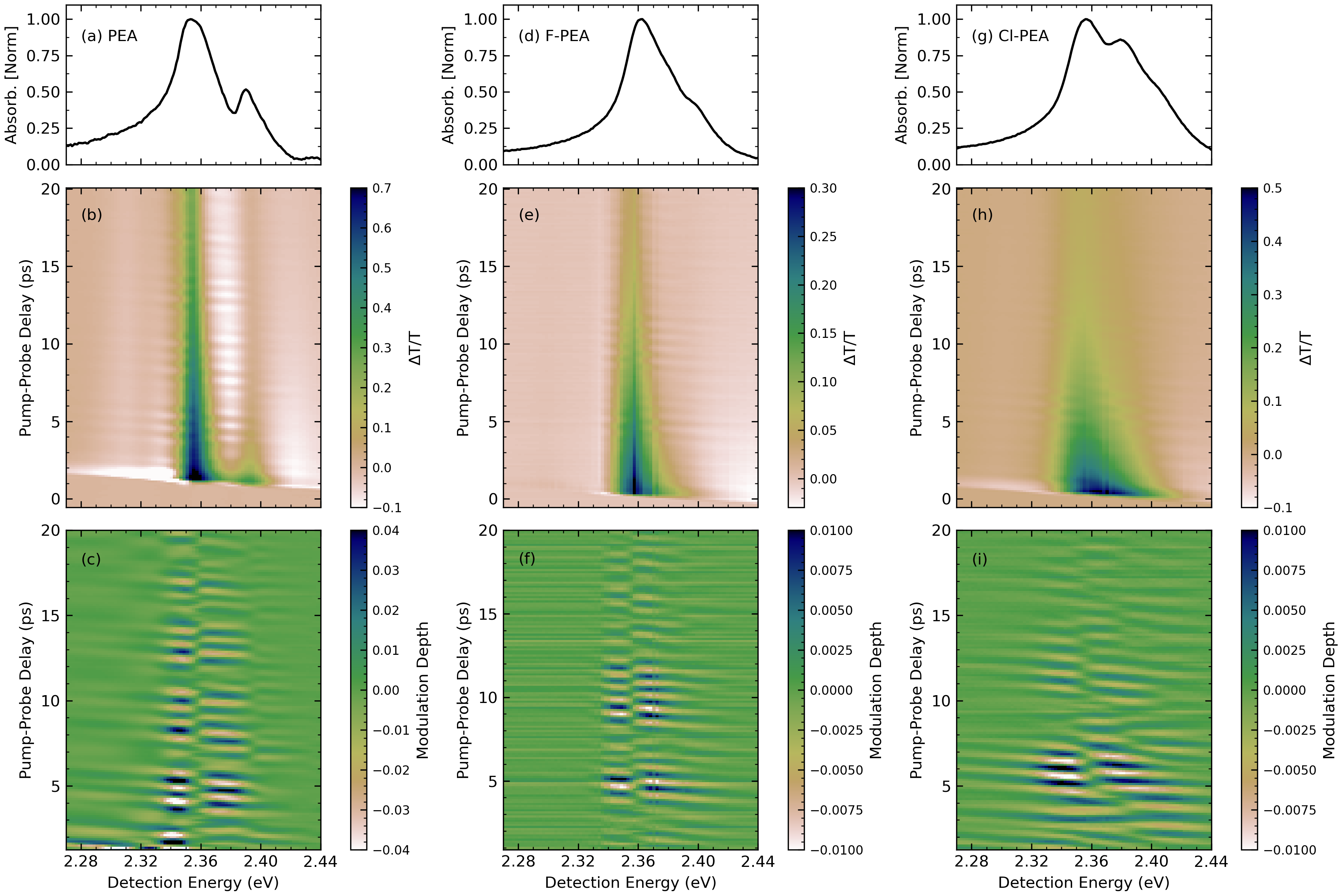}
    \caption{Linear absorption, transient absorption, and coherent phonon osciillations for \ce{(PEA)2PbI4} (a)-(c), \ce{(F-PEA)2PbI4} (d)-(f), and \ce{(Cl-PEA)2PbI4} (g)-(i) respectively. All measurements were taken at 15 K. }
    \label{fig:PP_oscillations}
\end{figure}

The electronic dynamics are subtracted from the differential transmission spectra using a high-order polynomial fit (see Fig. S\ref{fig:PP_oscillations}). The resulting modulated response is then fast Fourier transformed at each probe-energy to obtain a beating map. The Raman spectra shown in the text  are obtained by integrating over the relevant probe energies (see Fig. 3). 

\begin{figure}[h]
    \centering
    \includegraphics[width=6.5cm]{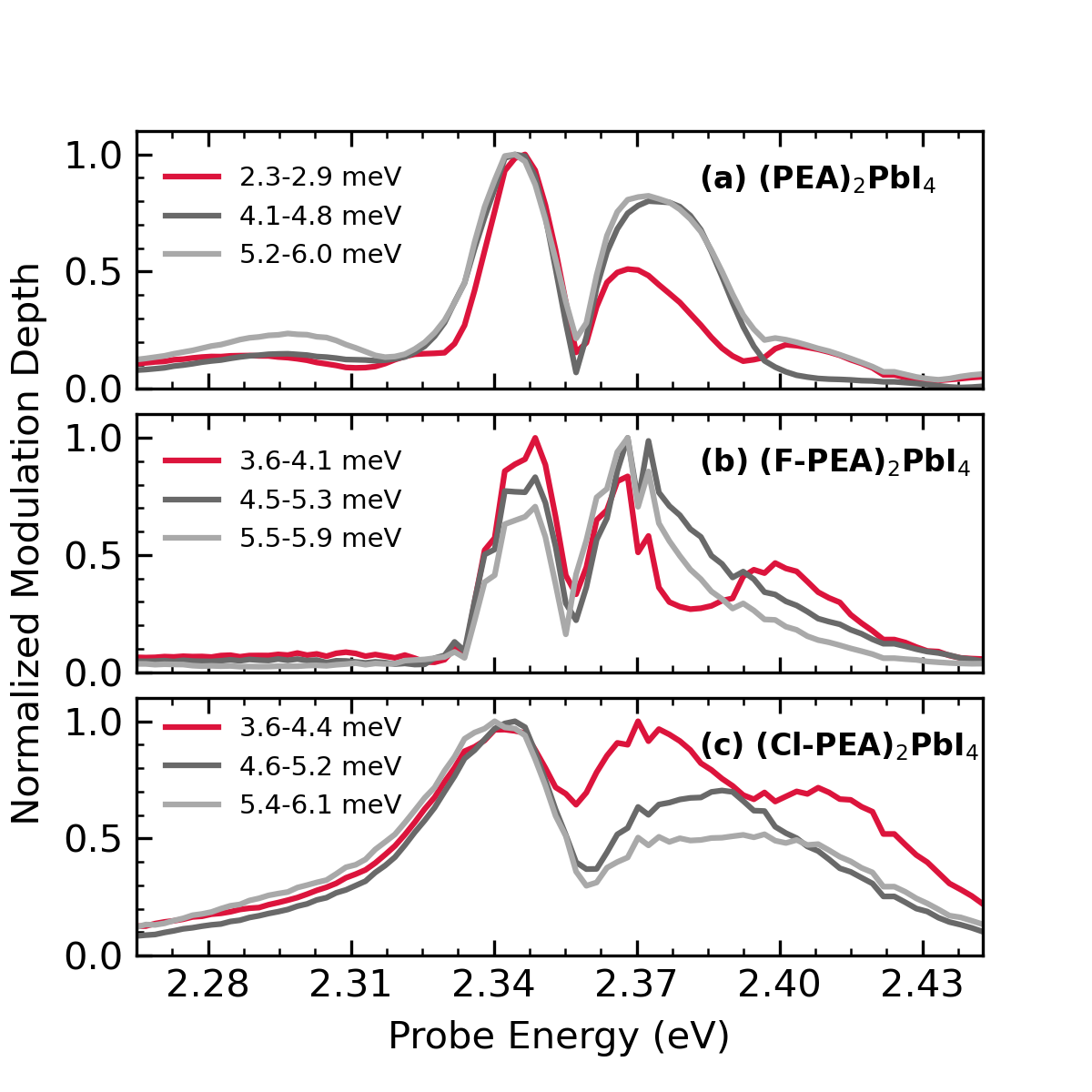}
    \caption{Spectral cuts of RISRS beating map at various phonon energies for \ce{(PEA)$_2$PbI$_4$} and its halogenated derivatives. All measurements were taken at 15 K.}
    \label{fig:raman_cuts}
\end{figure}
\subsection{Two-Dimensional Electronic Spectroscopy}
We employed 2D spectroscopy on \ce{(F-PEA)2PbI4} using the previously implemented scheme \cite{thouin2018stable, thouin2019enhanced}, developed and described in detail by Turner and coworkers \cite{turner2011invited}. The pulses used for these measurements were generated by a home-built single pass non-collinear optical parametric amplifier pumped by the third harmonic of a Yb:KGW ultrafast laser system (Pharos Model PH1-20-0200-02-10, Light Conversion) emitting 1030-nm pulses at 100kHz, with an output power of 20 W and pulse duration of 220-fs. The  pulses  were  individually  compressed  using  a  home-built  implementation  of  a  pulse  shaper  using  a  chirp  scan \cite{loriot2013self}.   The  resulting  pulse  duration  was 25 fs full-width at half-maximum (FWHM), as measured by second-harmonic generation cross-frequency-resolved optical gating (SHG-XFROG). All measurements were carried out in a vibration-free closed-cycle cryostat (Montana Instruments).
\subsection{1Q Rephasing Linewidth Analysis}
Four wave mixing spectroscopy measures coherent emission due to a third-order polarization induced by a sequence of three phase-locked femtosecond pulses in a "boxcar" geometry. The coherent emission is detected through spectral interferometry with a fourth attenuated pulse (local oscillator, LO) copropagating with the emitted field. We focus our analysis on the rephasing spectra where the emitted signal is acquired at $\Vec{k}_{sig} = -\Vec{k}_{a}+\Vec{k}_{b}+\Vec{k}_{c}$. This 2D lineshape allows for the separation of the homogeneous and inhomogenous broadening of the optical linewidths by simultaneously fitting the diagonal cut and antidiagonal cuts across the maximum of each diagonal features to the model described by the following equations \cite{siemens2010resonance, thouin2019enhanced}:

\begin{equation} \label{antidiag}
S_{AD}(\omega_{ad})=\left | \frac{
\text{exp}\left(\frac{(\gamma - i\omega_{ad})^2}{2\delta\omega^2}\right )
\text{erfc}\left (\frac{(\gamma - i\omega_{ad})}{\sqrt{2}\delta\omega}\right )}{\delta\omega(\gamma-i\omega_{ad})}
\right |
\end{equation}

and the diagonal lineshape,
\begin{multline} \label{diagonal}
S_{D}(\omega_{d}) = \sum_{j} \alpha_j \left|
\frac{\text{exp}\left(\frac{[\gamma - i(\omega_{d}-\omega_j)]^2}{2\delta\omega^2}\right)}{\gamma\delta\omega} \left[ \text{erfc}\left (\frac{[\gamma - i(\omega_{d}-\omega_j)]}{\sqrt{2}\delta\omega} \right)\right.\right. \\ + 
\left.\left.\text{exp}\left(\frac{2\gamma i(\omega_d-\omega_j)}{\delta\omega^2}\right)\text{erfc}\left(\frac{[\gamma + i(\omega_d-\omega_j)]}{\sqrt{2}\delta\omega}\right)\right] \right|
\end{multline}

\begin{figure}[h]
    \centering
    \includegraphics[width=10cm]{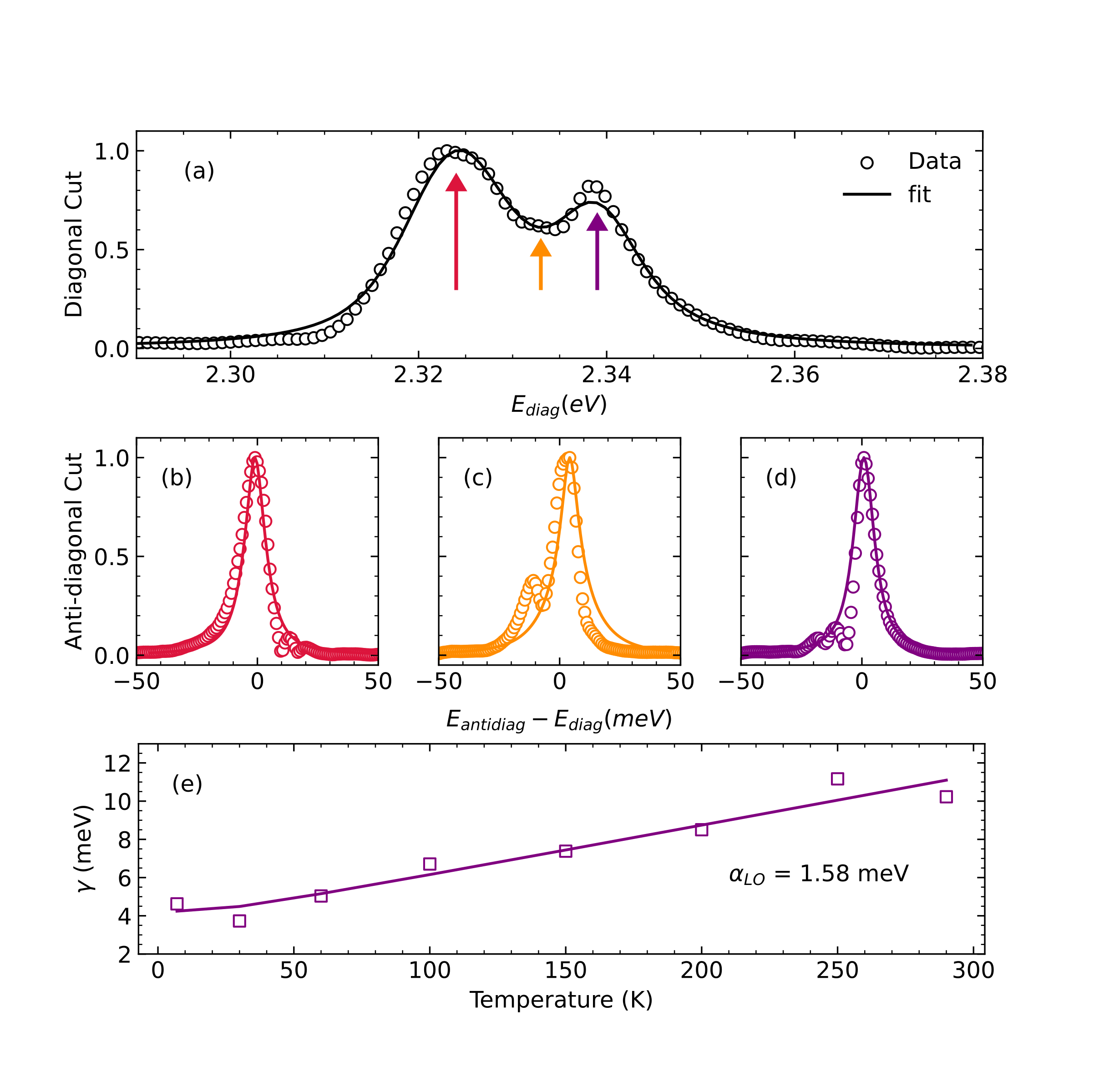}
    \caption{Fitting (a) diagonal and (b)-(c) antidiagonal spectral cuts of \ce{(F-PEA)2PbI4} to a lineshape model. The arrows in panel (a) mark the position along the diagonal where the antidiagonal cuts cross it. (e) Temperature dependence of the exciton dephasing rates for the exciton B ($X_B$) resonance. Squares represent the experimental linewidths and the line of best fit is shown for Eq. \ref{dephasing}.}
    \label{fig:dephasing}
\end{figure}

We measured 2D 1Q rephasing spectra at various sample temperatures to analyze exciton dephasing, $\gamma$, as a function of temperature. The measured $\gamma$ for the third diagonal feature corresponding to exciton B ($X_B$) is shown in Fig.~\ref{fig:dephasing} as a function of a temperature. The measured values are fit to:

\begin{centering}
\begin{equation} 
\label{dephasing}
\gamma(T) = \gamma_{T=0}+\alpha_{LO}\left[\frac{1}{\text{exp}(\text{E}_{LO}/k_BT)-1} \right]
\end{equation}
\end{centering}which assumes that line broadening arises from the scattering of excitons with a single thermally populated phonon mode with an effective energy E$_{LO}$ and an effective interaction parameter $\alpha_{LO}$. 

\end{document}